\documentstyle[twocolumn,aps]{revtex}
\input {epsf.tex}
\begin{document}
\draft
\title{
%\hfill\normalsize{astro-ph/9504097}\\
%\vspace{.2in}\\
\Large\bf Reconciling Inflation with Openness}
\author{Luca Amendola, Carlo Baccigalupi and
Franco Occhionero}

\address{
Osservatorio Astronomico di Roma,
Viale del Parco Mellini, 84, 00136 Roma, Italy
}
%\address{I--00136 Rome, Italy}
%\address{$^2$NASA/Fermilab Astrophysics Center, PO Box 500,
 %Batavia IL 60510,  USA}

%\date{\today}
\maketitle

\vspace{2.cm}
\begin{abstract}

It is already understood that
the increasing observational evidence for an open Universe may be
reconciled with inflation if our horizon is
contained inside one single huge bubble nucleated during the inflationary
phase transition. In the  scenario we present here, the Universe consists of
infinitely many superhorizon bubbles, like our own, the distribution of which
can be made to peak at $\Omega_0\approx 0.2$.
Therefore, unlike the existing literature, we do not have to rely upon
the anthropic principle nor upon special initial conditions.

\end{abstract}

%\vfill
%\centerline{To be published on Phys. Rev. D}
\vspace{.3cm}

%\newpage

	\section{\normalsize\bf Introduction}
	 An open Universe, with $\Omega_0\approx 0.2$, seems
	 to fit most astronomical observations. In connection with
	 CDM, it gives the best fit to the observed clustering (see e.g.
	 Ref. \cite{CFA2}); a similar value is requested for explaining the
	 dynamics of bound objects on relatively small scales (see e.g.
	 Ref. \cite{PEE}); it also increases the
	 age of the Universe, alleviating the conflict
	 with the age of the globular clusters. Further, $\Omega_0\approx 0.2$
	 is  in better agreement with direct geometrical estimates from
	 number counts
	 (see e.g. Ref. \cite{PEE} \cite{RP}).

However, as it is well known, inflation predicts $\Omega_0=1$.
This contradiction is one of the most interesting problems in
modern cosmology. It is certainly possible that the observations  in favour
of an open Universe are too limited to be representative of the whole
Universe, and that higher values of $\Omega_0$ will be found at larger scales.
On the other hand,
it is also possible to choose the initial conditions in
inflation so as to give $\Omega_0=0.2$ today, either by starting with
an extremely small density parameter at the beginning of
inflation, or by assuming that inflation lasted less than
60 $e$-foldings or so. Both possibilities, however,
introduce that fine-tuning of the initial conditions that
inflation itself tried to overcome;
moreover, there would be also a conflict with the
microwave background isotropy
\cite{KTF}.

Two ways out of the enigma have been proposed so far. The first, the single-
bubble scenario \cite{GW}
\cite{JAP} \cite{BGT}, assumes that a single giant bubble nucleated
from a false vacuum (FV) state, when the Universe was already flat
because of previous inflationary expansion,
 and inflated for about 60 $e$-foldings afterward:
our horizon is contained inside the bubble, and appears to be locally
open. The weak point of this model is the
old graceful exit problem:  inflation never ends
outside the bubble, and there is no reason why we should live inside
that infinitesimal fraction of space which nucleated out of the false vacuum,
unless one invokes anthropic arguments. Other problems of
the single-bubble model have
been discussed in Ref. \cite{LIN}.
In the second proposal, the many-bubbles scenario \cite{LIN}, one has
a two-field potential:  while a field drives the
inflationary slow-rolling, the second one performs
a phase transition, generating bubble-like open Universes, with
all possible density parameters, from zero to unity,
and all possible sizes. Here the
phase transition completes, all the Universe eventually nucleates
out of the FV state, but again there is no reason to expect
nor a preferential value of $\Omega_0$, nor
a bubble size big enough to contain
our horizon. Actually, it is difficult to
avoid the conclusion that most of the volume
is inside bubbles nucleated at the end
of inflation, thus exponentially smaller than our observed Universe.
Linde \cite{LIN} argues that eventually quantum cosmology will explain why
we live in a open Universe (if we really do).

The model we propose in this work implements a many-bubble scenario
in which the nucleation rate varies in such a way to give a $\Omega_0=0.2$
Universe with maximal probability. In our model, the peak of the
bubble nucleation can be chosen to occur early enough to have
super-horizon-sized bubbles, approximating a $\Omega_0=0.2$ Universe by today,
and narrow enough to consider our Universe as typical.
In other words, a flat, huge (or infinite) Universe appears to be
 composed of
locally open super-horizon-sized bubbles. It is remarkable that local
observations,  $\Omega_0$ and the amplitude of
perturbations, and the ``natural'' assumption
  that our Universe is typical, put strong constraints on
 the theory parameters, i.e. the
fundamental scales of the inflation and of the
primordial phase transition.

Our model, outlined in the next Section, has been already
introduced in Ref. \cite{OA} to produce large scale power
out of the remnants of the primordial phase transition. All what
we have to do here is to determine the parameters so to tune
the nucleation peak at super-horizon scales.

\section{\normalsize\bf The model}

To realize our scenario we need two pre-requisites.
First, we need two channels,
a false vacuum channel, to drive the
inflation
in the parent Universe, and a true vacuum channel, to drive
the shorter inflation inside the bubbles.
Second, we need a tunneling rate  tunable in time, so as to
produce a nucleation peak at the right time.
It is remarkable that the same model that we introduced in Ref. \cite{OA}
has just these features. This is certainly not the unique possibility,
but we will show that it is a rather simple one.

The model works in fourth order gravity \cite{S}, and
 exploits two fields: one, the scalaron $R$ (i.e. the Ricci
 scalar)
drives the slow-rolling inflation; the second, $\psi$, performs
the first-order phase transition. The phase transition
dynamics is governed both
by the potential of $\phi$ and by its coupling to $R$; the dynamics
of the slow-roll is ``built-in'' in the fourth order Lagrangian.
We already presented our model in detail in Ref. \cite{OA};
here we sketch its main features.
Our starting point is the Lagrangian density ${\cal L}=
{\cal L}_{grav}+{\cal L}_{mat}$, where
($c=\hbar=G=1$)
	\begin{equation} \label{lagra}
	{\cal L}_{grav}=-R+{R^2 \over  6 M^{2}W(\psi)}\;,
	\end{equation}
	and
  \begin{equation} \label{canonical}
       {\cal L}_{mat}=
	16\pi\left({1\over 2}\psi_{;\mu}\psi^{;\mu}-V(\psi)\right)\;.
  \end{equation}
  The coupling of the scalaron with $\psi$ can be thought of as
  a field-dependent effective mass
%\begin{equation} \label{reduced}
	$~		M_{eff}(\psi)=M W^{1/2}(\psi)\,,$
%\end{equation}
just like in Brans-Dicke gravity the coupling is a field-dependent
Planck mass. Putting
 $\alpha\equiv \log (a/a_{in})$,
 one finds in the slow-rolling regime
\begin{equation}\label{slowroll}
\alpha={ R-R_{in} \over 4 M_{eff}^2 } \,,
\end{equation}
which is essentially the only dynamical equation we need. In the
following, the instant labelled {\it in} will correspond to the
beginning of the last $N_T\approx 60$ $e$-foldings  of inflation.
The theory (\ref{lagra}) can be conformally transformed \cite{W} into
canonical gravity with the new metric
\begin{equation}\label{conformal}
\tilde g_{\alpha\beta} =e^{2\omega}g_{\alpha\beta}\,,\quad
e^{2\omega}=\left|{\partial {\cal L}\over \partial R}\right|
=1- { R \over 3M_{eff}^{2} }\;.
\end{equation}
 Once written
in the conformal frame, our model becomes
 indistinguishable from a
ordinary gravity theory with two fields governed by a specific
potential.

In the slow-rolling approximation useful relations link $\omega, H$ and
the number $N=N_T-\alpha$ of $e$-foldings to the end of inflation:
\begin{equation}\label{ef}
{4\over 3}N =(e^{2\omega}-1)={ 4 H^2 \over M_{eff}^2} \,.
\end{equation}
Correspondingly, $H_{in}=M_{eff}(N_T/3)^{1/2}$. The value $N_T$ is
fixed by the request that the largest observable scale, $L_h=2H_0^{-1}$,
was crossing out the horizon $N_T$ $e$-foldings before the end
of inflation, and it is close to 60 for standard
cosmological values \cite{KT}.
{}From (\ref{lagra}) and (\ref{conformal})  we obtain then Einstein gravity
with two scalar
fields $\psi$ and $\omega$, coupled by a potential given by
	\begin{equation}\label{conpot}
	U(\psi,\omega)
	=e^{-4\omega}\left[V(\psi)+{3M^2\over 32\pi}
	W(\psi)(1-e^{2\omega})^2  \right]\,.
	\end{equation}
The choice of a quartic for $W$ and a mass term for
  $V$  realizes the two
conditions discussed above:
\begin{equation}\label{ansatz}
	W(\psi)=1+{8\lambda\over \psi_0^4} \psi^2
        (	\psi-\psi_0)^2\,, \quad
		V(\psi)={1\over 2}m^{2}\psi^{2}\,.
	\end{equation}
This carves in fact in (\ref{conpot}) two parallel channels of different
height, separated by a peak  at $\psi_{PK}=\psi_0/2$.
The degeneracy of $W(\psi)$ in $\psi=0$ and $\psi=\psi_0$
is indeed removed by $V(\psi)$;
the true vacuum (TV) channel
remains at $\psi_{TV}=0$, while the false vacuum (FV) channel
is slightly displaced from  $\psi_0$.

In Ref. \cite{OA} we evaluated the tunneling rate $\Gamma$ for our model,
 defined as
\begin{equation}
\Gamma={\cal M}^4 \exp(-S_E)\,,
\end{equation}
where ${\cal M}$ is of the order of the energy of the false vacuum, and $S_E$
is the minimal Euclidean action, i.e. the action for the
so-called bounce solution of the Euclidean equation of motion.
The calculation of $S_E$
is simplified in the limit $\omega\gg 1$, i.e. for
$N\gg 1$. In this case in fact
\begin{equation}
U(\psi,\omega)\approx {3M^2\over 32 \pi} W(\psi)+V(\psi) e^{-4\omega} \,,
\end{equation}
and we can  directly use Coleman's formulas \cite{Coleman}
 to evaluate $S_E$,
provided we are in the  thin wall limit (TWL).
If $U_{FV}$ ($U_{TV}$) is the potential energy of the false vacuum
(true vacuum) state,
and $U_{PK}$ is the energy at the top of the barrier, the TWL
is guaranteed if $U_{PK}\gg U_{FV}-U_{TV}$,
i.e. if
\begin{equation}\label{twl}
\lambda\gg 8\pi {m^2\psi_0^2\over M^2}\,.
\end{equation}
 The result is (Ref.  \cite{OA})
\begin{equation} \label{sen1}
S_E=(N/N_1)^4 \,,
\qquad N_1^2={3\sqrt{3}\over 4}
{m^3\psi_0\over M^2\lambda}\,.
\end{equation}
The Eulidean action decreases as $N$ decreases;
 the bubble nucleation is then more likely
to occur during the last stages of inflation than at earlier times.
Finally, we can write the relevant parameter $Q=4\pi \Gamma/9H^4$, i.e. the
number of bubbles per horizon volume per Hubble time (quadrihorizon,
for short), as
	\begin{equation}\label{basic}
	Q(N)=	\exp  \frac{N_0^4-N^4}{N_1^4}\,,
	\quad
	 \left(\frac{{\cal M}} {M_{eff}}\right)^4=
\frac{9}{64 \pi} \exp \left(\frac{N_0}{N_1}\right)^4 \,
	\end{equation}
	where we have introduced a new parameter
	$N_0$  to mark the time at which there is on
	average one bubble per quadrihorizon,
	roughly corresponding to the end of the phase transition.
	It is useful to keep in mind that, as we will
	show later, $N_0\approx N_T\gg N_1$.
	$N_0$, or ${\cal M}$, in principle, can be derived in terms of the
	potential parameters \cite{Coleman}; the derivation is,
	however, very difficult: as customarily done in
	extended inflation (see e.g. \cite{LW}),
	 we will determine $N_0$ by requesting that
	 the transition eventually ends.

To summarize, our model has four characteristic quantities:
the slow-rolling inflationary rate, set by $M$;
the difference in energy between vacua states, set by $m$;
the barrier height, set by $\lambda$; and finally the separation between
the vacua states, set by $\psi_0$. These constants completely define
the slow-rolling and the phase transition dynamics.
In the next section we proceed to the evaluation of the
tunneling probability,
and show that we can tune the parameters to give $\Omega_0=0.2$ today with
maximal probability.

	\section{\normalsize\bf The tunneling function}

Let the number of bubbles per
horizon
nucleated in the time interval $dt$ be $dn_B$, where \cite{GW}
\begin{equation}
{dn_B\over dt}=
\Gamma a^3 V_{in} \exp \left[-{4\pi \over 3}
\int_{-\infty}^{t} dt'\Gamma(t')
\left(a(t')\int_{t'}^{t} {dt'' \over a(t'')}  \right)^3\right] \,,
\label{ourspe}
\end{equation}
 where $\Gamma=9H^4 Q/4\pi$ is the nucleation rate,
and where $a_0^3V_{in}=4\pi /3H_{in}^3$ is the horizon volume at
 $N_T$.
 The quantity $dn_B/dt$ is proportional to the tunneling
 rate per volume, $\Gamma$, and to the FV volume left at the time $t$
 (the volume not already occupied by bubbles).
 If after a certain time
 the  exponential term in (\ref{ourspe})
 decreases faster than $a^3$ increases,
 the FV volume fraction decreases; if this decrease
 is faster
 than $\Gamma$ increases, then $dn_B/dt$ will have a turnaround
 somewhere, indicating that the transition
 is being completed. The rest of this paper is essentially
 devoted
 to find the condition for this maximum to occur at the right time.

The $e$-folding time $N$ is defined as $N=N_T-\alpha$, where
$\alpha=\log(a/a_{in})=Ht$, so that we can put
\begin{equation}
\label{nt}
N=N_T(1-Ht/N_T)\,,\quad t(N_{T})=0\,.
\end{equation}
Since in all what follows we have $N\approx N_T$ (i.e., the
nucleation occurs around $N_T$), we can write
\begin{equation}\label{qq}
 \Gamma={9H^4\over 4\pi}\exp [r_0-r_T(1- t/\tau)]\,,
\end{equation}
where
\begin{equation}
r_0=(N_0/N_1)^4\,,\quad
r_T=(N_T/N_1)^4\,, \quad
\tau=N_T/4H\,.
\end{equation}
We will neglect the mild dependence of $H$ on $N$ compared to
the exponential dependence of $\Gamma$.
During inflation, $a(t)\approx a_0 \exp Ht$, so that we can
integrate easily
the  argument of the
exponential in eq. (\ref{ourspe}), and obtain finally
%\begin{equation}
%I=-B\exp{\left({tr_T\over \tau}\right)}\,,\end{equation}
\begin{equation}\label{dndt}
{dn_B\over  dt}=A \exp\left [t (3H+r_T/\tau) - B \exp(r_T t/\tau) \right]
\,,
\end{equation}
where
%\begin{equation}
$A=3He^{r_0-r_T}\,,$ and
%\label{A}\end{equation}
\begin{equation}\label{B}
B=3e^{r_0-r_T}{N_T\over 4r_T}g(N_T,N_1)\,,
\end{equation}
and where
\begin{eqnarray}
g(N_T,N_1)&=& \nonumber\\
\left(1-{3\over 1+N_T/4r_T}+
{3\over 1+N_T/2r_T}\right.&-&\left.
{1\over 1+3N_T/4r_T}\right)\,.
\end{eqnarray}
Expanding the argument of the exponential to the second order,
we can approximate (\ref{dndt}) as a Gaussian curve.
The second order term in (\ref{qq}) should be included as well; however,
for the values of interest of $N_0$ and $N_1$, it is irrelevant.
The result is
\begin{equation}\label{dn-t}
{dn_B\over  dt}=A'\exp\left[-{1\over 2} {(t-t_p)^2\over \sigma^2}\right]
\,,
\end{equation}
where
\begin{equation}
t_p={3H\tau^2+ r_T\tau (1-B)\over r_T^2 B }\,,\quad
\sigma^2=\tau^2/(r_T^2 B )\,,
\label{tp}
\end{equation}
and where the preexponential factor is
%\begin{equation}
$A'=A \exp(-B+ t_p^2/2\sigma^2)\,.$
%\end{equation}
The instant $t_p$ marks the peak of the nucleation
process; this will be fixed by the request to have
bubbles (slightly)  larger than the present horizon.
We must have nucleation before $N_T$, or simply $t_p<0$,
and it is clear that a necessary condition is in any case
$B>0$.
%Approximating  again $N_0\approx N_T$,  we can put
%$t_p=-[3(N_0-N_T)-1]e^{(r_T-r_0)}/3H$:
%then in order to have $t_p<0$ (nucleation before $N_T$),
%we must  have simply
%\begin{equation}\label{tpneg}
%3(N_0-N_T)>1\,.
%\end{equation}
%put $B\gg 1$, so that
%\begin{equation}
%t_p= - \tau / r_T\,,\qquad \sigma^2=t_p^2/B
%\end{equation}
%(and $A'=A\exp(-B/2)$).

It is now useful to use $N$ as time variable. It follows then that
\begin{equation}\label{dn-n}
\left|{dn_B\over  dN}\right|=(A'/H) \exp\left[-{1\over 2} {(N-N_p)^2\over
\sigma_N^2}\right]\,,
\end{equation}
with
\begin{equation}
%N_p=N_T\left(1+e^{(r_T-r_0)}{[3(N_0-N_T)-1]\over3N_T}\right)\,,\qquad
%\sigma_N^2=e^{r_T-r_0} {N_T\over 12r_T}={N_{T}^{2}\over 16r_{T}^{2}B}\,.
N_p=N_T\left(1-{Ht_p\over N_T}\right)\,,\quad
\sigma_N^2={N_{T}^{2}\over 16r_{T}^{2}B}\,,
\end{equation}
where we require that $N_p>N_T$.
%This shows that the approximations we have adopted are consistent:
%if $N_0\approx N_T$ then also $N_p\approx N_T$; and if
%$N_T\gg N_1$, then the Gaussian
%is narrow enough ($\sigma_N^2\ll N_T^2$ )
%to approximate $N$ around $N_T$.
We need now the relation between the present curvature
$C_0=|\Omega_0-1|$ and the nucleation time $N$.
We know that if a bubble nucleates at $N$ its density
parameter is
\begin{equation}
\label{ome}
\Omega={U_{TV}\over U_{FV}}=
[1+\gamma N^{-2}]^{-1}\approx 1-\gamma N^{-2}\,,
\end{equation}
where the constant
%\begin{equation}
$~\gamma\equiv 3\pi m^2\psi_0^2/M^2~$
%\end{equation}
has been taken small (compared to $N^2$) for simplicity of calculation,
even if it is not necessary. Now,
from the Friedmann equation, in the limit
of small deviation from flatness, it is easy to see that
 the relation between the curvature today and
 the curvature during inflation
is $C\approx N^{-1} e^{2(N-N_T)} C_0$.
However, when the curvature deviates sensibly from zero, this relation is no
longer acceptable. Neglecting the RDE phase, it is possible to derive
from the standard solution of an open Universe (see e.g. \cite{KT})
the following expression
\begin{equation}
C(N)\approx N^{-1} e^{2(N-N_T)} C_0f(C_0)\,,
\end{equation}
where
%\begin{equation}
$~f(C_0)=1+C_0^{3/2}/(1-C_0)\,.$
%\end{equation}
Thus, we get finally
\begin{equation}\label{finally}
C_0f(C_0)=\gamma e^{2(N_T-N)} /N\approx  \gamma e^{2(N_T-N)} /N_T\,.
\end{equation}
It appears then that, as we would expect,
$C_0\to 0$ if the bubble nucleates at
$N\to\infty$, and $C_0\to 1$ if the nucleation occurs at small $N$.
Only a  value $N\approx N_T$ would result in intermediate values
of $C_0$.
{}From Eq. (\ref{finally}), we can express the nucleation rate
directly in terms of the present curvature $C_0$:
\begin{equation}
{dn_B\over  dC_0}={A'\over 2H}\left({1\over C_0}
+{f'\over f}\right)\exp\left\{-{1\over 2}
{\left
[\log \left({C_0f(C_0)\over C_pf(C_p)}\right)\right]^2\over \sigma_C^2}
\right\}\,,
\end{equation}
with $f'\equiv df(C_0)/dC_0\,,$
%\begin{equation}
$~C_p f(C_p)=\gamma e^{2Ht_p}/N_T\,,$ and
%\qquad
$~\sigma_C^2=4\sigma_N^2\,.$
%\end{equation}
We can then define the probability
$P(\Omega_0)$ that we live  in a $\Omega_0$
Universe as
\begin{equation}\label{dn-c}
P(\Omega_0)\equiv
{dn_B\over  dC_0} {L(C_0)^3\over L_h^3}\,,
\end{equation}
where $L(C_0)$ is the comoving size
of a bubble which today has curvature $C_0$. As we will see shortly,
$P(\Omega_0)$ is normalized to unity.
Eq. (\ref{dn-c}) is the central result of this work;
it gives the probability density to live today  in a bubble with
given $\Omega_0$. As we have seen, we have obtained it using the
standard tools of inflationary cosmology,
without employing
arguments from quantum cosmology. Most importantly,
we have obtained $P(\Omega_0)$ {\it without assuming
special initial conditions}.

We now impose three
observational constraints on $P(\Omega_0)$:
first, we must impose the condition that the bubbles fill the Universe,
i.e. that the transition completes; second,
we require the nucleation peak to occur for $\Omega_0=0.2$,
the present observational
value; and third, we require a narrow distribution, say
$\sigma_C \ll 1$, to ensure our Universe is typical.
This implies three conditions on our
parameters $N_0, N_1$ and $\gamma$.
The first constraint amounts to
 requiring that the volume contained
in  bubbles
nucleated from $N=\infty$ down to $N=0$ be equal to, or larger than,
 the present horizon:
\begin{equation}
\label{norm}
\int_{\infty}^0 {dn_B\over dN} L^3 dN\ge  L_h^3\,.
\end{equation}
where $L(N)$ is the comoving scale of a bubble nucleated
at $N$.
In terms of $P(\Omega_0)$, Eq. (\ref{norm})
 is simply the condition that
the probability be normalized to unity (or to a value larger than unity,
which simply indicates that the bubbles are too densely packed to
assume spherical symmetry).
Since
\begin{equation}
H(N)L(N)\approx H_{in}L_h \exp (N-N_T)\,,
\end{equation}
we have the normalization condition
\begin{equation}\label{normcond}
\int_{\infty}^0 {dn_B\over dN } e^{3(N-N_T)} dN\ge  1\,,
\end{equation}
neglecting again the time dependence of $H$ during inflation.
{}Using directly the form (\ref{dndt}) where we express
$t$ in terms of $N$ with (\ref{nt}) it follows
%\begin{equation}
%A'/H\ge {1\over\sqrt{2\pi}\sigma_N}
%e^{-{9\over 2}\sigma_N^2+3(N_{T}-N_{p})}\,.
%\end{equation}
\begin{equation}\label{cond}
{3N_Te^{r_0-r_T}\over 4r_TB}\left(1-e^{-Be^{4r_T}}\right)
\approx  3g(N_T,N_1)^{-1}     \ge 1\,.
\end{equation}
We display in Fig. 1 the region of the plane ($N_1,N_0$)
which satisfies the constraint above (with $N_T=60$),
for which the curves (\ref{dn-n}) are sharply peaked
around $N_{p}$ $(\sigma_N^2\ll 1)$, and
for which $N_p>N_T$; as one can see, there is a vast
region in which our theory is successful.
Finally, the condition that $C_p=0.8$ gives
\begin{equation}
\label{c-cond}
\gamma\approx 4N_T e^{-2Ht_p}\,.
\end{equation}
{}From Fig. 1 one sees that
 $N_1$ has to be roughly larger than $10$, and
 therefore
 $\gamma\approx 10^3$ (from (\ref{tp}) with $N_0\ge 60$) and, by
 the condition
(\ref{sen1}),
\begin{equation}
 (\sqrt{3} m\gamma /4\pi \psi_0 \lambda)^{1/2}\ge 10\,.
\end{equation}
Let us summarize the constraints to which the  parameters
are subject. The tunneling function has to be peaked at $N$
slightly larger than $N_T$; the value of $\Omega$ at that time
has to be such that today $\Omega_0\approx 0.2$;
the peak should be narrow, so that the probability
to live in a $\Omega_0=0.2$ Universe is high; and the
transition should be intense enough to fill the Universe with bubbles.
A further condition is that the slow roll do not generate too strong
inhomogeneities; roughly, this implies $M< 10^{-5}$ in Planck units
\cite{OA}. Finally, we have to be in  the thin wall limit
(\ref{twl}).
It is remarkable that our model meets easily all these requirements.
For instance, we can have  $N_1\approx 25$ and $N_0-N_T\approx 2$,
so that $\gamma\approx 10^{3}$ (from (\ref{c-cond})); then, fixing
$M=10^{-6}$ and $\lambda =10^4$ (respecting (\ref{twl})), we have
$m\approx 1$ (the Planck scale), $\psi_0\approx 10^{-5}$
(near the scale of $M$).
In Fig. 2 we plot $P(\Omega_0)$ for this set
of parameters.

\section{\normalsize\bf Conclusions}

We presented a scenario in which the flat, inflationary Universe is
filled  by super-horizon-sized underdense bubbles, which
approximate open Universes. This reconcile the astronomical observations
in favour of $\Omega_0=0.2$ with inflation. Our own bubble-Universe is one
of an infinite number of similar bubbles. Our model differs
from  the  single-bubble scenario \cite{GW}
\cite{JAP} \cite{BGT} in which one must invoke anthropic
arguments to explain our position, and is different also from the
model presented by Linde \cite{LIN}, in which there is no reason to
expect preferential nucleation for any given value
of $\Omega$. As we showed in the previous Section, we
can tune the parameters to achieve maximal probability for the
nucleation of
$\Omega_0=0.2$ bubbles, without assuming special
initial conditions,
and satisfying all other
constraints.

It is worth remarking again that the
measure of $\Omega_0$ along with the assumption that the Universe had
an
inflationary epoch, and that our position is typical, put
strong constraints on the fundamental parameters of
the primordial potential.
It is interesting to observe that the
phase transition parameters $\lambda,\psi_0,m$ would be
unobservable either if the nucleation occurred much earlier, because
then the subsequent expansion would have again flattened the space, or
much later, because then the very small sub-horizon bubbles would have
thermalized, recovering again a $\Omega=1$ Universe.

Inside the bubbles one has the usual mechanism of generation of
inflationary perturbations
\cite{RP} \cite{JAP} \cite{BGT}. In Ref. \cite{OA} we presented  a scenario
in which extra power is provided by the bubble-like structure of
a primordial phase transition. It is possible that reducing the
local $\Omega_0$ to $0.2$ is enough to reconcile canonical CDM with
large scale structure, so that no further phase transitions
need to be invoked. However, evidences are increasing toward the
presence of huge voids in the distribution of matter in the
present Universe, and for velocity fields that are difficult to
explain without a new source of strong inhomogeneities.
If this is the case, the possibility of an additional
primordial phase transition occurred around 50 $e$-foldings
before the end of inflation
should be seriously taken into account.

% \end{document}

 \newpage
% \null
 \onecolumn
 \begin{figure}
%\vspace{1cm}
%\epsfysize=4cm
%\epsfbox{/kosu/luca/figures/openf1.ps}
%\epsfbox{openf1.ps}
%\epsfysize=0.5pt
\epsfbox[0 0 30 550]{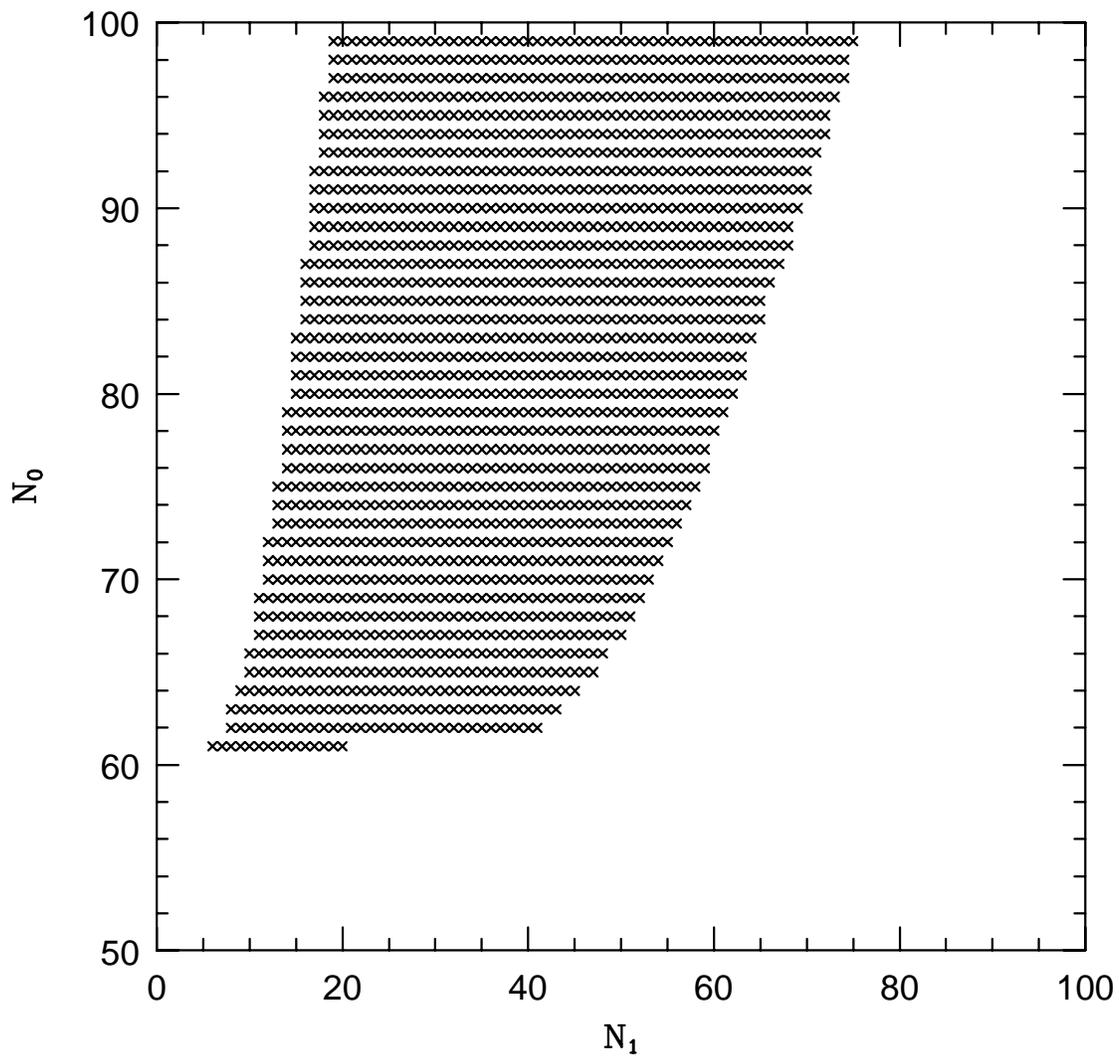}
\caption{The shaded region indicate the values for $N_1$ and $N_0$ for which
  our theory is successful:  underdense
  bubbles with a narrow peaked distribution
containing our horizon's scale today fill the whole Universe.}
\label{fig1}
\end{figure}
\newpage
\begin{figure}
%\vspace{5cm}
%\epsfysize=4cm
%\epsfbox{/kosu/luca/figures/openf2.ps}
%\epsfbox{openf2.ps}
%\epsfysize=0.5pt
\epsfbox[0 0 30 550]{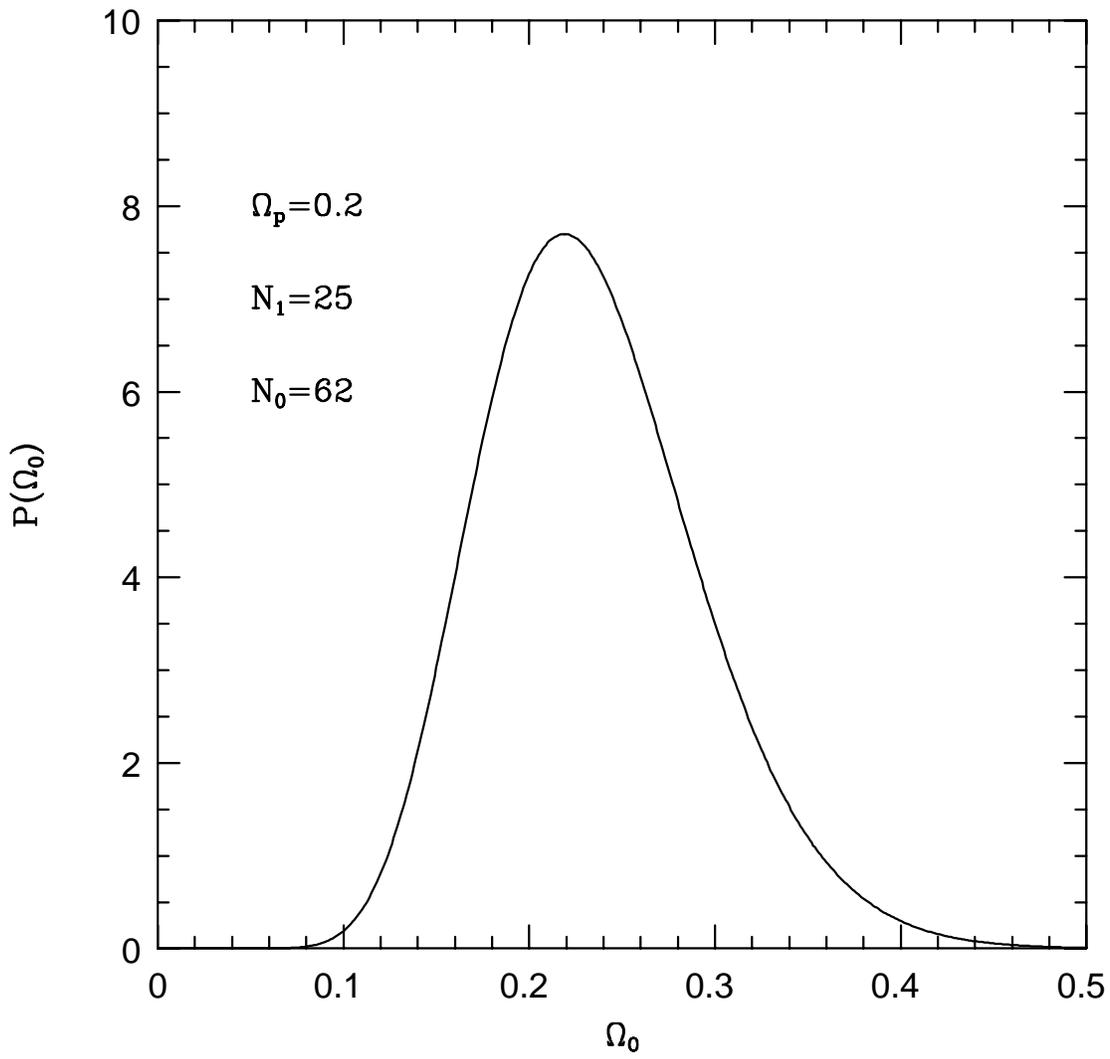}
\caption{Plot of $P(\Omega_0)$ for $N_0=62$ and $N_1=25$.
Our present curvature is the most probable, and the distribution is
very narrow around its mean value.}
\label{fig2}
\end{figure}
%\epsfbox{openf1.ps}
%\epsfbox{openf2.ps}

\end{document}